\documentclass[]{aastex631}

\shorttitle{BINGO Telescope - status update}
\shortauthors{C.A.Wuensche et al.}
\graphicspath{{./}{figures/}}
\begin{document}

\title{Baryon Acoustic Oscillations from Integrated Neutral Gas Observations: an instrument to observe the 21cm hydrogen line in the redshift range 0.13 $<$ z $<$ 0.45 - status update}

\correspondingauthor{C.A.Wuensche}
\email{ca.wuensche@inpe.br}

\author[0000-0003-1373-4719]{Carlos. A. Wuensche}
\affiliation{Divis\~ao de Astrof\'{i}sica, Instituto Nacional de Pesquisas Espaciais.  Av. dos Astronautas 1758, Jd. da Granja, CEP 12227-010, S\~ao Jos\'e dos Campos - SP, Brazil}

\author[0000-0002-9814-1690]{Elcio Abdalla}
\affiliation{Instituto de F\'{i}sica, Universidade de S\~ao Paulo, R. do Mat\~ao, 1371 - Butant\~a, S\~ao Paulo - SP, 05508-09, Brazil.}

\author[0000-0003-2063-4345]{Filipe B. Abdalla}
\affiliation{Department of Physics and Electronics, Rhodes University, PO Box 94, Grahamstown, 6140, South Africa}
\affiliation{Divis\~ao de Astrof\'{i}sica, Instituto Nacional de Pesquisas Espaciais. Av. dos Astronautas 1758, Jd. da Granja, CEP 12227-010, S\~ao Jos\'e dos Campos - SP, Brazil}
\affiliation{Instituto de F\'{i}sica, Universidade de S\~ao Paulo, R. do Mat\~ao, 1371 - Butant\~a, S\~ao Paulo - SP, 05508-09, Brazil.}
\affiliation{University College London, Gower Street, London,WC1E 6BT, UK}

\author[0000-0002-5455-2682]{Luciano Barosi}
\affiliation{Unidade Acad\^emica de F\'{i}sica, Universidade Federal de Campina Grande, R. Apr\'{i}gio Veloso, 58429-900 - Bodocong\'o, Campina Grande - PB, Brazil}

\author[0000-0003-3419-9153]{Bin Wang}
\affiliation{School of Aeronautics and Astronautics, Shanghai Jiao Tong University, Shanghai 200240, China}
\affiliation{Center for Gravitation and Cosmology, College of Physical Science and Technology, Yangzhou University, 225009, Yangzhou, China}

\author[0000-0003-4931-6105]{Rui An}
\affiliation{Department of Physics and Astronomy, University of Southern California, Los Angeles, CA, 90089, USA}
\affiliation{School of Aeronautics and Astronautics, Shanghai Jiao Tong University, Shanghai 200240, China}

\author[0000-0002-3111-1331]{João A. M. Barreto}
\affiliation{Instituto de F\'{i}sica, Universidade de S\~ao Paulo, R. do Mat\~ao, 1371 - Butant\~a, S\~ao Paulo - SP, 05508-09, Brazil. }

\author[0000-0002-4269-515X]{Richard Battye}
\affiliation{Jodrell Bank Centre for Astrophysics, Department of Physics and Astronomy, The University of Manchester, Oxford Road, Manchester, M13 9PL, U.K.}

\author[0000-0001-9465-6868]{Francisco A. Brito}
\affiliation{Departamento de Física, Universidade Federal da Paraíba, Caixa Postal 5008, 58051-970 João Pessoa, Paraíba, Brazil}
\affiliation{Unidade Acad\^emica de F\'{i}sica, Universidade Federal de Campina Grande, R. Apr\'{i}gio Veloso, 58429-900 - Bodocong\'o, Campina Grande - PB, Brazil}

\author[0000-0002-0743-8430]{Ian Browne}
\affiliation{Jodrell Bank Centre for Astrophysics, Department of Physics and Astronomy, The University of Manchester, Oxford Road, Manchester, M13 9PL, U.K.}

\author[0000-0002-8733-8535]{Daniel Correia} 
\affiliation{Instituto de F\'{i}sica, Universidade de S\~ao Paulo, R. do Mat\~ao, 1371 - Butant\~a, S\~ao Paulo - SP, 05508-09, Brazil.}

\author[0000-0002-3670-9826]{André A. Costa}
\affiliation{Center for Gravitation and Cosmology, College of Physical Science and Technology, Yangzhou University, 225009, Yangzhou, China}

\author[0000-0002-7217-4689]{Jacques Delabrouille}
\altaffiliation[10]{IRFU, CEA, Universit\'e Paris Saclay, 91191 Gif-sur-Yvette, France}
\affiliation{Department of Astronomy, School of Physical Sciences, University of Science and Technology of China, Hefei, Anhui 230026}
\affiliation{Laboratoire Astroparticule et Cosmologie (APC), CNRS/IN2P3, Universit\'e Paris Diderot, 75205 Paris Cedex 13, France}

\author[0000-0002-0045-442X]{Clive Dickinson}
\affiliation{Jodrell Bank Centre for Astrophysics, Department of Physics and Astronomy, The University of Manchester, Oxford Road, Manchester, M13 9PL, U.K.}

\author[0000-0001-7438-5896]{Chang Feng}
\affiliation{School of Astronomy and Space Science, University of Science and Technology of China, Hefei, 230026, China}

\author[0000-0002-5032-8368]{Elisa G. M. Ferreira}
\affiliation{Max-Planck-Institut fr Astrophysik, Karl-Schwarzschild-Str. 1, 85748 Garching, Germany}

\author[0000-0003-0578-9533]{Karin Fornazier}
\affiliation{University College London, Gower Street, London,WC1E 6BT, UK}
\affiliation{Instituto de F\'{i}sica, Universidade de S\~ao Paulo, R. do Mat\~ao, 1371 - Butant\~a, S\~ao Paulo - SP, 05508-09, Brazil.}

\author[0000-0003-2899-2171]{Giancarlo de~Gasperis}
\affiliation{INFN Sez. di Roma 2, Via della Ricerca Scientifica, 1 I-00133 Roma, Italy}
\affiliation{Dipartimento di Fisica, Universit\`a degli Studi di Roma Tor Vergata, Via della Ricerca Scientifica, 1 I-00133 Roma, Italy}

\author[0000-0001-9997-8313]{Priscila Gutierrez}
\affiliation{Instituto de F\'{i}sica, Universidade de S\~ao Paulo, R. do Mat\~ao, 1371 - Butant\~a, S\~ao Paulo - SP, 05508-09, Brazil.}

\author[0000-0001-7911-5553]{Stuart Harper}
\affiliation{Jodrell Bank Centre for Astrophysics, Department of Physics and Astronomy, The University of Manchester, Oxford Road, Manchester, M13 9PL, U.K.}

\author[0000-0001-6801-3519]{Ricardo G. Landim}
\affiliation{Technische Universit\"at M\"unchen, Physik-Department, James-Franck-Strasse 1, 85748 Garching, Germany}

\author[0000-0001-5506-5125]{Vincenzo Liccardo}
\affiliation{Jodrell Bank Centre for Astrophysics, Department of Physics and Astronomy, The University of Manchester, Oxford Road, Manchester, M13 9PL, U.K.}
\affiliation{Divis\~ao de Astrof\'{i}sica, Instituto Nacional de Pesquisas Espaciais.  Av. dos Astronautas 1758, Jd. da Granja, CEP 12227-010, S\~ao Jos\'e dos Campos - SP, Brazil} 

\author[0000-0001-8108-0986]{Yin-Zhe Ma}
\affiliation{NAOC-UKZN Computational Astrophysics Centre (NUCAC), University of KwaZulu-Natal, Durban, 4000, South Africa}
\affiliation{School of Chemistry and Physics, University of KwaZulu-Natal, Westville Campus, Private Bag X54001, Durban 4000, South Africa}

\author[0000-0002-6631-1245]{Telmo Machado}
\affiliation{Divis\~ao de Astrof\'{i}sica, Instituto Nacional de Pesquisas Espaciais.  Av. dos Astronautas 1758, Jd. da Granja, CEP 12227-010, S\~ao Jos\'e dos Campos - SP, Brazil}

\author[0000-0002-2214-5329]{Bruno Maffei}
\affiliation{IAS, Université Paris-Saclay, 91405 Orsay Cedex, France}

\author[0000-0002-6519-6038]{Alessandro Marins}
\affiliation{Instituto de F\'{i}sica, Universidade de S\~ao Paulo, R. do Mat\~ao, 1371 - Butant\~a, S\~ao Paulo - SP, 05508-09, Brazil.}

\author[0000-0002-4646-4762]{Milena M. M. Mendes} 
\affiliation{Instituto de F\'{i}sica, Universidade de S\~ao Paulo, R. do Mat\~ao, 1371 - Butant\~a, S\~ao Paulo - SP, 05508-09, Brazil.}

\author[0000-0002-4672-5467]{Eduardo Mericia}
\affiliation{Divis\~ao de Astrof\'{i}sica, Instituto Nacional de Pesquisas Espaciais.  Av. dos Astronautas 1758, Jd. da Granja, CEP 12227-010, S\~ao Jos\'e dos Campos - SP, Brazil}

\author[0000-0002-3178-363X]{Christian Monstein}
\affiliation{ETH Zürich, Institute for Particle Physics and Astrophysics, HIT J13.2, Wolfgang-Pauli-Strasse 27, 8093 Zürich, Switzerland}

\author[0000-0002-3416-6258]{Pablo Motta}
\affiliation{Instituto de F\'{i}sica, Universidade de S\~ao Paulo, R. do Mat\~ao, 1371 - Butant\~a, S\~ao Paulo - SP, 05508-09, Brazil.}

\author[0000-0002-5845-3649]{Camila Novaes}
\affiliation{Divis\~ao de Astrof\'{i}sica, Instituto Nacional de Pesquisas Espaciais.  Av. dos Astronautas 1758, Jd. da Granja, CEP 12227-010, S\~ao Jos\'e dos Campos - SP, Brazil}

\author[0000-0001-7482-262X]{Carlos H. N. Otobone}
\affiliation{Instituto de F\'{i}sica, Universidade de S\~ao Paulo, R. do Mat\~ao, 1371 - Butant\~a, S\~ao Paulo - SP, 05508-09, Brazil.}

\author[0000-0003-3412-2586]{Michael Peel}
\affiliation{Instituto de Astrof\'{i}sica de Canarias, 38200 La Laguna, Tenerife, Canary Islands, Spain}
\affiliation{Departamento de Astrof\'{i}sica, Universidad de La Laguna (ULL), 38206 La Laguna, Tenerife, Spain}

\author[0000-0002-4785-5589]{Am\'{i}lcar R. Queiroz}
\affiliation{Unidade Acad\^emica de F\'{i}sica, Universidade Federal de Campina Grande, R. Apr\'{i}gio Veloso, 58429-900 - Bodocong\'o, Campina Grande - PB, Brazil}

\author[0000-0001-8338-6356]{Christopher Radcliffe}
\affiliation{Phase2 Microwave Ltd., Unit 1a, Boulton Rd, Pin Green Ind. Est., Stevenage, SG1 4QX, UK}

\author[0000-0001-9126-6266]{Mathieu Remazeilles}
\affiliation{Jodrell Bank Centre for Astrophysics, Department of Physics and Astronomy, The University of Manchester, Oxford Road, Manchester, M13 9PL, U.K.}

\author[0000-0003-2903-0679]{Rafael M. G. Ribeiro}
\affiliation{Instituto de F\'{i}sica, Universidade de S\~ao Paulo, R. do Mat\~ao, 1371 - Butant\~a, S\~ao Paulo - SP, 05508-09, Brazil.}

\author[0000-0002-2631-7781]{Yu Sang}
\affiliation{Center for Gravitation and Cosmology, College of Physical Science and Technology, Yangzhou University, 225009, Yangzhou, China}

\author[0000-0002-5549-5747]{Juliana F. R. Santos}
\affiliation{Instituto de F\'{i}sica, Universidade de S\~ao Paulo, R. do Mat\~ao, 1371 - Butant\~a, S\~ao Paulo - SP, 05508-09, Brazil.}

\author[0000-0001-7015-998X]{Larissa Santos}
\affiliation{Center for Gravitation and Cosmology, College of Physical Science and Technology, Yangzhou University, 225009, Yangzhou, China}

\author[0000-0002-1408-6947]{Marcelo V. Santos}
\affiliation{Unidade Acad\^emica de F\'{i}sica, Universidade Federal de Campina Grande, R. Apr\'{i}gio Veloso, 58429-900 - Bodocong\'o, Campina Grande - PB, Brazil}

\author[0000-0002-7114-8139]{Chenxi Shan}
\affiliation{Center for Gravitation and Cosmology, College of Physical Science and Technology, Yangzhou University, 225009, Yangzhou, China}

\author[0000-0002-4746-2013]{Gustavo B. Silva}
\affiliation{Instituto de F\'{i}sica, Universidade de S\~ao Paulo, R. do Mat\~ao, 1371 - Butant\~a, S\~ao Paulo - SP, 05508-09, Brazil.}

\author[0000-0003-3839-9814]{Frederico Vieira}
\affiliation{Divis\~ao de Astrof\'{i}sica, Instituto Nacional de Pesquisas Espaciais.  Av. dos Astronautas 1758, Jd. da Granja, CEP 12227-010, S\~ao Jos\'e dos Campos - SP, Brazil}

\author[0000-0001-8283-5295]{Jordany Vieira}
\affiliation{Instituto de F\'{i}sica, Universidade de S\~ao Paulo, R. do Mat\~ao, 1371 - Butant\~a, S\~ao Paulo - SP, 05508-09, Brazil.}

\author[0000-0002-4284-8988]{Thyrso Villela}
\affiliation{Instituto de F\'{i}sica, Universidade de Bras\'{i}lia, Bras\'{i}lia, DF, Brazil}
\affiliation{Divis\~ao de Astrof\'{i}sica, Instituto Nacional de Pesquisas Espaciais.  Av. dos Astronautas 1758, Jd. da Granja, CEP 12227-010, S\~ao Jos\'e dos Campos - SP, Brazil}

\author[0000-0003-3198-3591]{Linfeng Xiao}
\affiliation{School of Aeronautics and Astronautics, Shanghai Jiao Tong University, Shanghai 200240, China}

\author[0000-0002-6486-6765]{Weiqiang  Yang}
\affiliation{Liao Ning Normal University, No.850 Huanghe Road Shahekou District, Dalian,Liaoning, P.R.China 116029}

\author[0000-0002-4117-343X]{Jiajun  Zhang}
\affiliation{CTPU-IBS, Theory building, 4th Floor, 55, Expo-ro, Yuseong-gu, Daejeon, Korea}

\author[0000-0002-5840-9349]{Xue Zhang}
\affiliation{Center for Gravitation and Cosmology, College of Physical Science and Technology, Yangzhou University, 225009, Yangzhou, China}

\author[0000-0001-8443-6095]{Zenghao Zhu}
\affiliation{School of Physics and Astronomy, Shanghai Jiao Tong University, Shanghai 200240, P. R. China}

\collaboration{51}{(BINGO Collaboration)}

%% Note that the \and command from previous versions of AASTeX is now
%% depreciated in this version as it is no longer necessary. AASTeX 
%% automatically takes care of all commas and "and"s between authors names.

%% AASTeX 6.31 has the new \collaboration and \nocollaboration commands to
%% provide the collaboration status of a group of authors. These commands 
%% can be used either before or after the list of corresponding authors. The
%% argument for \collaboration is the collaboration identifier. Authors are
%% encouraged to surround collaboration identifiers with ()s. The 
%% \nocollaboration command takes no argument and exists to indicate that
%% the nearby authors are not part of surrounding collaborations.

%% Mark off the abstract in the ``abstract'' environment. 
\begin{abstract}

BINGO (\textbf{B}AO from \textbf{I}ntegrated \textbf{N}eutral \textbf{G}as \textbf{O}bservations) is a unique radio telescope designed to map the intensity of neutral hydrogen distribution at cosmological distances, making  the first detection of Baryon Acoustic Oscillations (BAO) in the frequency band 980 MHz - 1260 MHz, corresponding to a redshift range $0.127 < z <  0.449$. BAO is  one of the most powerful probes of cosmological parameters and BINGO was designed to detect the BAO signal to a level that makes it possible to put new constraints on the equation of state of dark energy. The telescope will be built in Para\'iba, Brazil and consists of two $\thicksim$ 40m mirrors, a feedhorn array of 28 horns, and no moving parts, working as a drift-scan instrument. It will cover a $15^{\circ}$ declination strip centered at $\sim \delta=-15^{\circ}$, mapping $\sim 5400$ square degrees in the sky. The BINGO consortium is led by University of S\~ao Paulo with co-leadership at National Institute for Space Research and Campina Grande Federal University (Brazil). Telescope subsystems have already been fabricated and tested, and the dish and structure fabrication are expected to start in late 2020, as well as the road and terrain preparation. 

\end{abstract}

%% Keywords should appear after the \end{abstract} command. 
%% The AAS Journals now uses Unified Astronomy Thesaurus concepts:
%% https://astrothesaurus.org
%% You will be asked to selected these concepts during the submission process
%% but this old "keyword" functionality is maintained in case authors want
%% to include these concepts in their preprints.

%% From the front matter, we move on to the body of the paper.
%% Sections are demarcated by \section and \subsection, respectively.
%% Observe the use of the LaTeX \label
%% command after the \subsection to give a symbolic KEY to the
%% subsection for cross-referencing in a \ref command.
%% You can use LaTeX's \ref and \label commands to keep track of
%% cross-references to sections, equations, tables, and figures.
%% That way, if you change the order of any elements, LaTeX will
%% automatically renumber them.
%%
%% We recommend that authors also use the natbib \citep
%% and \citet commands to identify citations.  The citations are
%% tied to the reference list via symbolic KEYs. The KEY corresponds
%% to the KEY in the \bibitem in the reference list below. 

\section{Introduction}
\label{sec:intro}

One of the main cosmological challenges in the 21st century is to explain the present acceleration in the expansion of the universe, first unequivocally inferred in 1998 by two independent groups measuring supernovae of type Ia, one led by \cite{Perlmutter:1998} and the other by \cite{Riess:1998}. This led to the 2011 Nobel Prize in Physics and can be explained by postulating a negative pressure from a new component, the so-called Dark Energy (hereafter, DE). In combination with other observations, such as the Cosmic Microwave Background (CMB), there is little doubt about the existence of such an accelerated phase and the main focus of observational cosmology is now to try to determine its detailed properties. 

Among the various programs to measure those properties, Baryonic Acoustic Oscillations (BAO) are recognised as one of the most powerful probes of the properties of DE (e.g., \cite{Albrecht:2006}; \cite{Weinberg:2013}). BAO are a signature in the matter distribution from the recombination epoch and are imprinted in the distribution of galaxies at redshifts less than that of the epoch of reionization ($z \lesssim 0.5$). In the context of the standard cosmological model ($\Lambda\textrm{CDM}$ model), BAO can be identified as a small but detectable excess of galaxies with separations of order of $150~\textrm{Mpc.h}^{-1}$ (\cite{BAO:Eisenstein2005}; \cite{BAO:Anderson2014}; \cite{Planck2018:cosmoparams}). This excess is the imprint of the acoustic oscillations generated during CMB times and its linear scale is known from basic physics. Consequently, a measure of its angular scale can be used to determine the distance up to a given redshift. The same oscillations produce the familiar  peaks observed in the CMB anisotropy power spectrum. 

To date, BAOs have only been detected by performing large galaxy redshift surveys in the optical waveband, where they are used as tracers of the underlying total matter distribution. It is important they are confirmed in other wavebands and measured over a wide range of redshifts. Since optical surveys demand more time to detect individual galaxies and cover a relatively small fraction of the sky, just a few percent of the local volume has been mapped with such a type of technique. The radio band provides a complementary observational window for studying BAO, and the natural radio tracer is the redshifted 21cm emission line. However, the volume emissivity associated with this line is low, meaning that detecting individual galaxies up to $z \sim 0.5$ requires a very substantial collecting area. It has been proposed that mapping the Universe measuring the collective 21cm emission of the underlying matter distribution, using a technique known as intensity mapping, hereafter IM \citep{Loeb:2008, Masui:2013, Switzer:2013, Peterson:2006}, can be many times more efficient than optical surveys (e.g., \cite{Chang:2007, Loeb:2008}). 

The BINGO (\textbf{B}AO from \textbf{I}ntegrated \textbf{N}eutral \textbf{G}as \textbf{O}bservations) telescope is a new instrument designed specifically for observing BAO in frequency band around $1~{\rm GHz}$ and to provide a new insight into the Universe at $ 0.13 \lesssim z \lesssim 0.45$ with a dedicated instrument. The optical configuration consists of a compact, two 40m diameter, static dishes with an exceptionally wide field-of-view ($ \sim 14.75^{\circ} \times 6.0^{\circ} $) and 28 feed horns in the focal plane. BAO observations will be carried through IM HI surveys (see, e.g., \cite{Peterson:2006}). BINGO's main scientific goal is to claim the first BAO detection in radio in this redshift range, mapping the three dimensional distribution of HI, yielding a fundamental contribution to the study of DE, with observations spanning several years. No detection of BAOs in radio has been claimed so far, making BINGO an interesting and appealing instrument in the years to come. Moreover, in view of the observation strategy and with some adjustment in its digital backend, BINGO will also be capable to detect transient phenomena at very short time scales ($\lesssim 1~{\rm ms}$), such as pulsars and  Fast Radio Bursts (FRB, \cite{Lorimer:2007,Cordes2019fast}).

Earlier versions of the BINGO telescope concept are described in \cite{Battye:2013, Battye:2016, Wuensche2018} and a comprehensive analysis of foregrounds and $1/f$ noise that can limit the BINGO performance was published by \cite{BigotSazy:2015}. The Project Design Review that occurred in July 2019 fixed the telescope location in the hills of ``Serra do Urubu'' in the municipality of Aguiar, Para\'{i}ba, Brazil (Lon: $38^{\circ}~16'~4.8''$ W, Lat: $7^{\circ}~2'~27,6'' $S). It presents a very clean radio environment, as reported by \cite{Peel:2019}, which is one of the most critical requirements for high-quality measurements. 

The main institutions in the BINGO consortium are:  University of S\~ao Paulo (USP), National Institute for Space Research (INPE) and Campina Grande Federal University (UFCG), in Brazil, University of Manchester and University College London (United Kingdom), and Yang Zhou University (China). The BINGO project was presented in the BRICS Astronomy Working Group 2019 and this paper contains the status of the project as of July 2020. 

\section{Science Goals}

Recent observational results have been useful in shedding new light onto aspects of dark matter and DE, which comprise about 95\% of the mass of the Universe, and it is expected that maybe a new physical picture emerges once we find more about the structure of the Dark Sector \citep{Abdalla:2009wr}. BAO studies are one of the most powerful probes of the properties of DE (see, e.g., \cite{Albrecht:2006, Weinberg:2013})  and have only been probed by optical galaxy redshift surveys   (see, e.g., \cite{BAO:Eisenstein2005,BAO:Percival2010,BAO:Tojeiro2014,BAO:Anderson2014,BAO:Reid2015}). 

Constraints on various existing DE models (see, e.g., \citep{Wang:2005,Wang:2007,Feng:2008,Micheletti:2009pk,He:2009,He:2011,Abdalla:2014cla}) coming from BAO measurements break the degeneracies between cosmological parameters that are present when inferring them from the CMB observations. This potentially allows to constrain extensions of the standard cosmological model which are usually degenerate with the Hubble constant.

Following the first detection of BAOs using SDSS surveys \cite{BAO:Eisenstein2005}, BAOs have been measured with several new generation galaxy and quasar redshift surveys. In particular, the recent measurements with the BOSS survey achieve a precision in the BAO scale of a few percent in the redshift range $z = [0, 2]$ \citep{BAO:Anderson2014,BAO:Tojeiro2014,BAO:Reid2015}. New large experiments are being designed or are already under construction to refine BAO measurements in the optical band, such as DESI \citep{Aghamousa:2016a,Aghamousa:2016b}, Euclid \citep{EUCLID:Refregier2010} and WFIRST \citep{Levi:2011,Green:2011}, as well as LSST for imaging \citep{TysonAPS:2002}. Their goal is to achieve a precision below the percent level at higher redshifts. Nevertheless, all of them are large endeavours and the first light of any of them is not expected before the mid-2020's decade.

This is why it is so important that the current results are confirmed in other wavebands and measured over a wide range of redshifts. The radio band provides a unique and complementary observational window for the understanding of DE via measurements of integrated redshifted 21cm hydrogen emission line  \citep{Abdalla:2009wr}. HI redshift data allow us to construct a unique three dimensional map of the mass distribution, with a different perspective compared to the data obtained with optical telescopes. Indeed, HI is expected to be one of the best tracers of the total matter \citep{Padmanabhan:2015}. 

\begin{figure}[h!]
\begin{center}
\includegraphics[width=10cm]{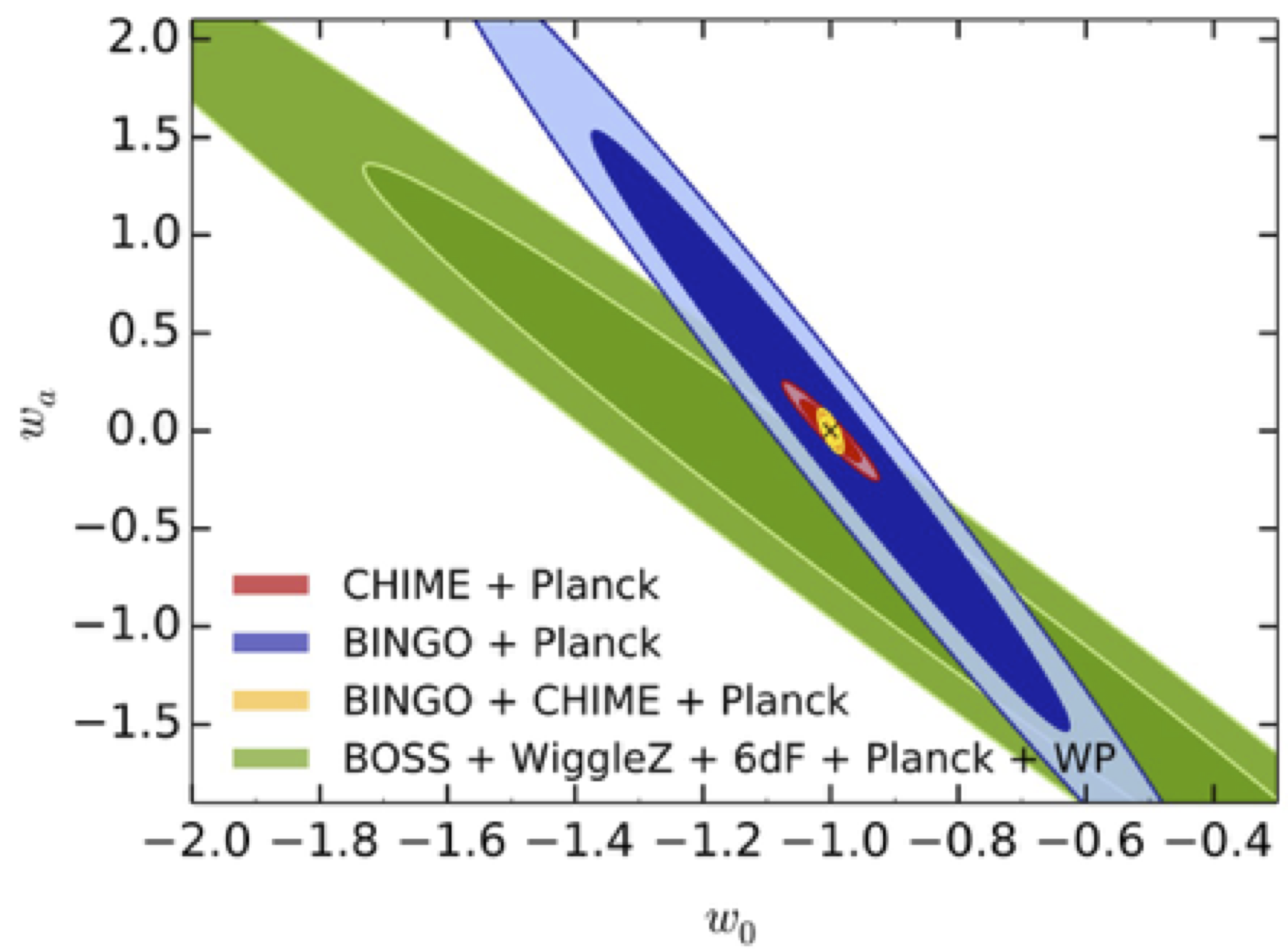}
\caption{Constraints on DE equation of state, including forecasts from CHIME and BINGO. Note the tight error elipse obtained with combined radio data (BAO from CHIME and BINGO + Planck).}
\label{BAOconstraints1}
\end{center}
\end{figure}

The combination of BAO radio measurements with other data sets (for instance, CMB and optical BAO) can be a powerful strategy to put more stringent constraints in the cosmological parameters analysis. Fig.~\ref{BAOconstraints1} shows the forecast of the constraints from BINGO, combined with other data sets, for the time dependent parametrisation the equation of state of the DE with $w_{0}$ and $w_a$ ($1^{st}$ time derivative of $w_{0}$) for various data sets. The constraints were computed using the Fisher matrix approach as described by \cite{Bull:2015}. 

The plot shows the improvement in the constraints obtained with the combination of simulated results from BINGO and CHIME \citep{CHIME:Newburgh2014} , compared to the current constraints given by the combination of the results of  the Planck and WMAP satellites \citep{Planck2013:cosmoparams,   Planck2016:overview,  Planck2018:cosmoparams,  WMAP:Bennett2013}, with the optical galaxy surveys (BOSS \citep{BAO:Anderson2014,  BAO:Tojeiro2014}, WiggleZ \citep{Kazin:2014} and 6dF \citep{Beutler:2011}). For constant $w$, the measurement would lead to an accuracy on $w$ of $\lesssim 8\%$ when including BINGO together with CHIME and Planck. 

Tight constraints can be obtained with radio data alone (BAO from CHIME and BINGO plus Planck), providing a completely independent measurement from optical surveys. The improvement from the red to the yellow contour is due to having two independent measurements at different redshifts, showing the importance of BINGO. We have made estimates of the projected errors of the DE parameters assuming that all the other cosmological parameters are fixed. BINGO will be competitive and complementary to optical surveys, which are limited by different systematic errors.

\begin{figure}[h]
\begin{center}
\includegraphics[scale=0.6]{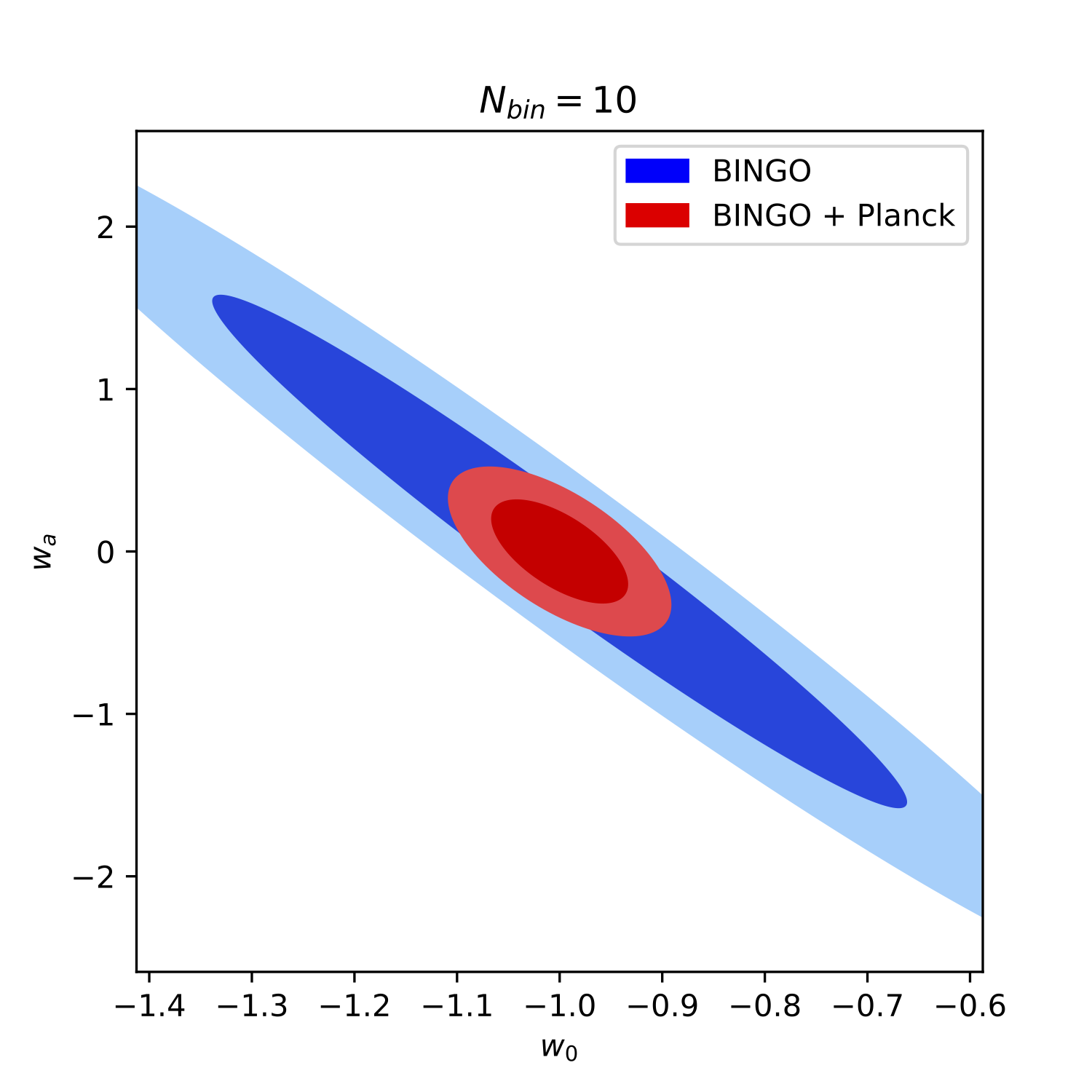}
\includegraphics[scale=0.6]{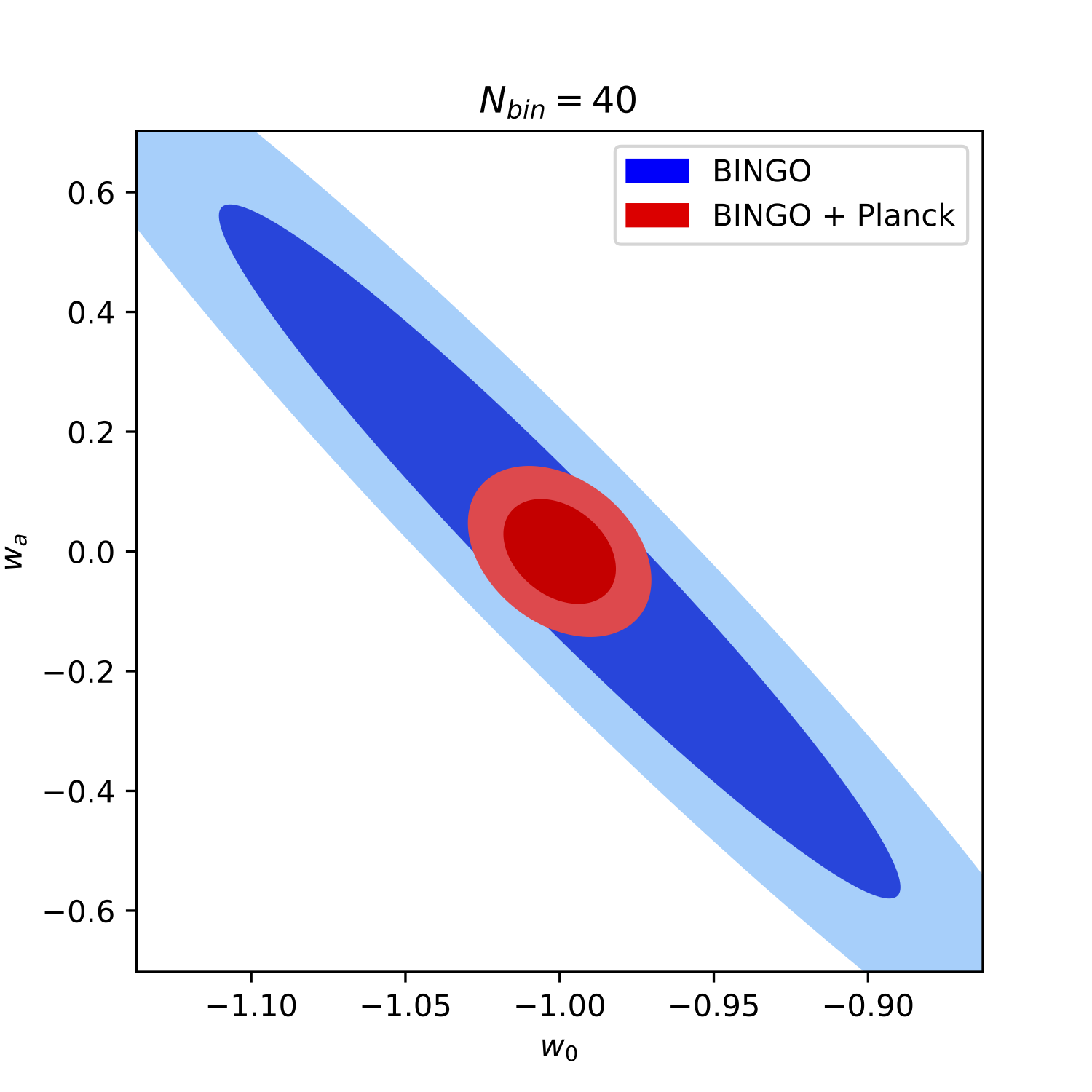}
\caption{\label{plot_w_wa} $68\%$ and $95\%$ confidence level constraints on the DE equation of state parameters using BINGO and BINGO + \textit{Planck}. The results are show for two different number of redshift bins \citep{BINGO_Costa:2020}.}
\label{BAOconstraints2}
\end{center}
\end{figure}

The combination of BINGO and Planck only also place good constraints in the values of the equation of state of DE.  In figure \ref{BAOconstraints2} we compare the $68 \%$ and $95 \%$ confidence levels constraints on the parameters of the DE equation of state using BINGO alone and BINGO plus Planck. The constraints improve with increasing number of redshift bins. Those figures can be compared with the optimal case results from \citep{Olivari:2018}. Our marginalised $68 \%$ c.l. constraints on the DE equation of state parameters are $2.4 \%$ for $w_0$ and $\sigma_{w_a} = 0.11$. Olivari et al. obtained $4 \%$ for a constant equation of state model. 

A more detailed work discussing the forecast methods under development for the BINGO data analysis is presented in \cite{BINGO_Costa:2020}.

\subsection{HI intensity mapping}

Together with other experiments, IM radio surveys can become a very promising complementary route to study BAOs. Indeed, optical experiments are required to gather data with very high angular resolution (of the order of 1 arcsecond) to detect individual galaxies, while the BAO scale is typically of the order of a degree. On the other hand, radio BAO experiments have angular resolutions which are well matched to the BAO scale and thus offer a very low cost-effective approach. IM experiments are a more direct tracer of dark matter being less sensitive to ``peak statistics'' bias. Moreover, radio experiments can use other elements/molecules (CO, $H-{\alpha}$) besides neutral hydrogen \citep{Kovetz:2017}, and are subject to different astrophysical effects and systematics when compared to optical surveys. The combination of their results with those obtained from optical surveys can significantly reduce systematic effects during cosmological parameter analysis. 

BINGO will also use IM to map the three-dimensional distribution of HI, with observations spanning a few years. Also, an experiment such as BINGO will pave the way to the extension of the use of 21cm IM  to high redshifts, opening a new window on cosmological epochs, which cannot be probed in the optical or near-infrared bands. Extensive simulations of the expected BAO signal and how it can be extracted are currently in progress, with results to be published in the near future. 

At frequencies below a few GHz the future of radio astronomy will be dominated by SKA (Square Kilometre Array) project \citep{SKA:Maartens2014}, that should begin operating by mid-2020's. Its design has been strongly driven to probe the HI 21cm line from its rest frequency (1420 MHz) down to 50 MHz. The great importance of the IM technique is highlighted by the decision to change the baseline design of the SKA Phase 1 to enable a IM approach  \citep{SKAWG:2020}. Albeit SKA is designed  to work mainly in interferometric mode, the proposed IM observations are intended to work in autocorrelation mode; i.e. using each dish separately, so that SKA and BINGO operation modes will be somewhat similar. Each BINGO beam will be equivalent to a single SKA dish, so that the hardware and software know-how acquired during BINGO development will be readily available to take advantage of the much larger signal to noise ratio achieved by SKA \citep{Bull:2015,Abdalla:2015}.

\section{The instrument}

Interferometers are the most suitable instruments to perform IM in the  angular resolution of $\sim 0.5^{\circ}$, covering very large areas of the sky. However, they also require expensive hardware and sophisticated electronics and time-keeping systems to make the necessary correlations. A number of approaches has been proposed to conduct IM surveys using interferometer arrays rather than a single dish (see, for instance, the papers from \cite{vanBemmel:2012cj, Pober:2012zz, Baker:2011jy}) and a detailed discussion about advantages and disadvantages for both was published by \cite{Bull:2015}. 

\begin{figure}[h]
\begin{center}
\includegraphics[width=12cm]{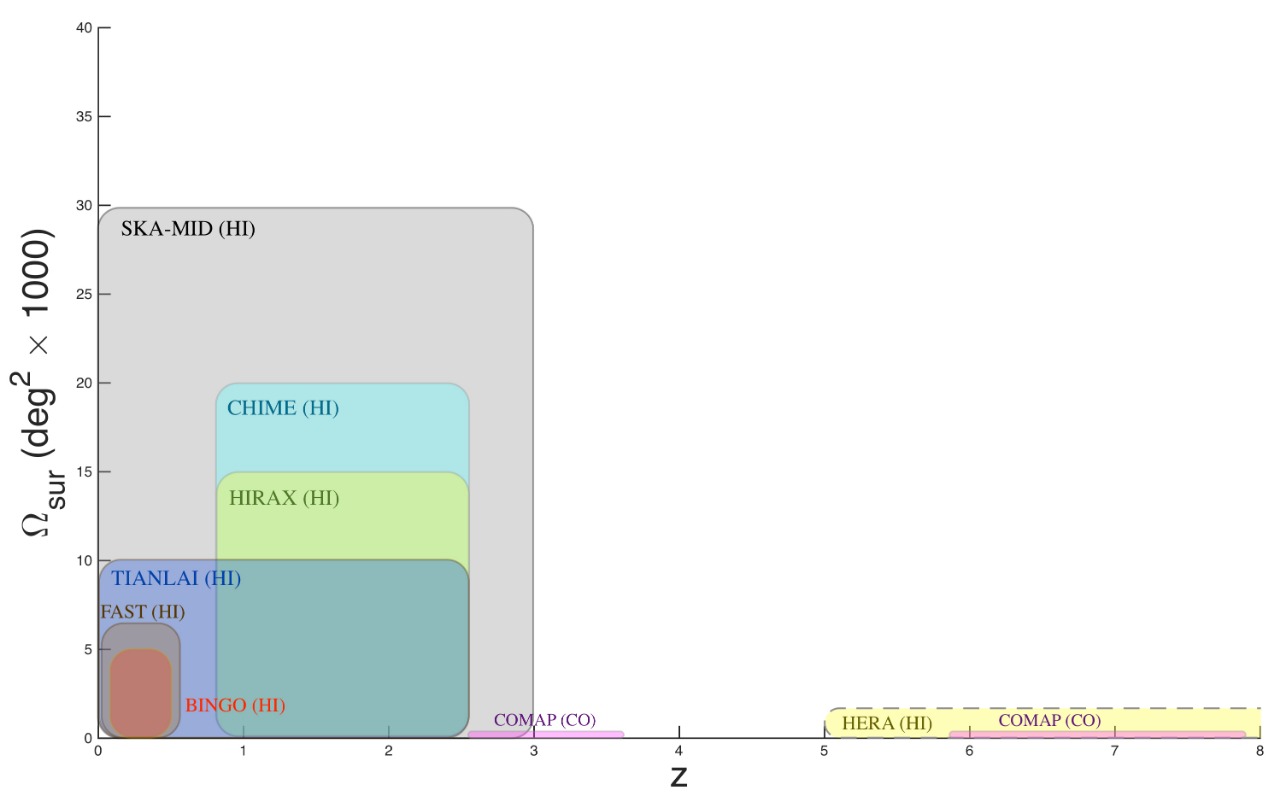}
\caption{Comparative plot of sky and redshift coverage for different IM experiments. CHIME, FAST, Tianlai and COMAP are in operation.}
\label{fig:surveys}
\end{center}
\end{figure}

Single-dish telescopes with stable receivers can be a low-cost, good efficiency approach to BAO \emph{IM} studies at redshifts ($z \lesssim 0.5$) \cite{Battye:2012} and BINGO belongs to this class.  %No detection of BAOs in radio has been claimed so far, making BINGO an interesting and appealing instrument in the years to come, with a strong potential for a pioneering discovery of the BAO in this wavelength. 
For comparison, we show some of the IM experiments planned for the near future, or already in operation, in figure\ref{fig:surveys}, with covered area in the sky as a function of redshift. Albeit operating in the same redshift range of  TIANLAI, FAST and SKA telescopes, we believe BINGO can be still be a strong competitor, producing good quality data before their relatives. 

The difficulty in doing a statistically significant BAO detection is due to the high intensity of the Galactic and extragalactic signals in the BINGO frequency band ($980 \le \nu \le 1260 {\rm GHz}$ ) . The HI signal is typically $\thicksim 200~\mu K$ whereas the foreground continuum emission from the Galaxy is $\thicksim 1~K$ with spatial fluctuations $\thicksim 100~mK$. Detecting signals of $\thicksim 200~\mu K$ with a non-cryogenic receiver implies that every pixel in our intensity map requires an accumulated integration time larger than 1 day over the course of the observing campaign, which is expected to last a few years. The total integration time can be built up by many returns to the same patch of sky but between these returns the receiver gains need to be highly alike and achieving this stability is a major design concern. 

Fortunately, the integrated 21cm emission exhibits characteristic variations as a function of frequency whereas the continuum emission has a very smooth spectrum; this allows for the two signals to be separated \citep{Chang:2007,Liu:2011hh,Chang:2010,Olivari:2016}. However, while there is a clear-cut statistical signature that allows for the separation of different components, the design of an instrument to measure the HI contribution to the detected signal needs to be very careful. It should contemplate a very clean and symmetric beam, with low sidelobe levels and very good polarisation purity, to avoid systematic effects that can result in leakage of the Galactic foreground emission, which is partially linearly polarised and concentrated towards the Galactic plane, into the HI signal. 

A declination strip of $\thicksim 15^{\circ}$, centered at $\delta \thicksim -15^{\circ}$ aims at minimising the Galactic foreground contamination and will be the optimal choice for the BINGO survey. The need to clearly resolve structures of angular sizes corresponding to a linear scale of around $150~Mpc$ at BINGO's chosen redshift range implies that the required angular resolution has to be $\thicksim 40'$. 

The general concept for the BINGO instrument is described in  \cite{Battye:2013, Battye:2016} and updated in \cite{Wuensche2018}. The current stage of the instrument construction is as follows: 1) horn prototype is completed, tested and approved for electromagnetic and mechanical performance; 2) front end, including polariser, transitions, OMT (``magic-tees'') and rectangular-to-coaxial transitions are completed, tested and approved for electromagnetic and mechanical performance; 3) the LNA and receiver chain is completed, passed the initial testing and needs to be improved for a better system temperature; 4) different LNA types are under testing to improve the first stage performance; 5) optical design is nearly completed, with the final dish parameters and focal plane arrangement simulated and tested in the mission pipeline data reduction software. The main parameters describing the telescope can be found in Table \ref{tab:optics}. 

\begin{table*}[ht]
 \scriptsize
 \centering
 \caption{BINGO telescope parameter list}
 \label{tab:optics}
 \begin{tabular}{|c|c|}
 \hline
System temperature (K) & 70  \\
Frequency band (MHz) & 980.00 - 1260.00  \\
Sampling time (Hz) & 10   \\
Instrument noise (mK, 1 second) & 23  \\
Polarizations & 2 \\
Redshift channels    & 30    \\
Redshift interval & 0.127 - 0.449  \\
Focal length (m) & 63.2   \\
Primary reflector F/D ratio		& 140   	\\
Instantaneous focal plane area (sqr. deg.) & $14.75 \times 6.0$  \\
 \hline
\multicolumn{2}{|c|}{\textbf{Primary mirror (parabola}} \\
Major semi-axis (m) & 25.7   \\
Minor semi-axis (m) & 20.0  \\
 \hline
\multicolumn{2}{|c|}{\textbf{Secondary mirror (hyperbola)}} \\
Major semi-axis (m) & 18.3   \\
Minor semi-axis (m) & 18.0   \\
 \hline
Pixel solid angle (sqr deg) & 0.504  \\
Optics FWHM (deg @ 1 GHz) & 0.67   \\
Survey area (square deg) & 5323   \\
Total \# horns (Phase 1) & 28  \\
\hline
\end{tabular}
\end{table*}

\subsection{The optics}

\begin{figure}[h]
\begin{center}
\includegraphics[width=7.5cm]{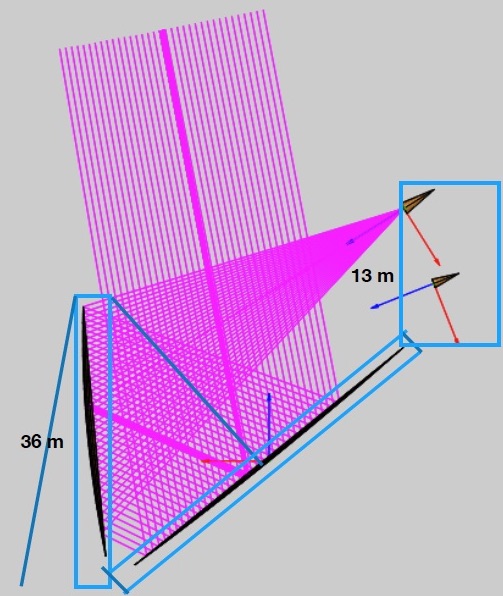}
\caption{Optical design schematics as presented in the July 2019 CDR \citep{BINGO_Abdalla:2020}.}
\label{fig:optical_design}
\end{center}
\end{figure}

The final optical system consists of an off-axis dual mirror telescope following the Dragone-Mizuguchi condition.
% in order to minimise the cross-polarisation. 
Combined with the use of corrugated feedhorns, similarly to CMB experiments, such configuration allows for low sidelobes, low spillover and superb polarisation performance. The focal plane includes 28 feed horns giving an overall field of view of $14.75^{\circ} \times 6^{\circ}$. The forward gain of each beam formed by the combination of the feedhorn and the telescope do not vary by more than 1 dB for all the pixels across the whole focal plane, while keeping low beam aberration and ellipticity. The current mirror and focal plane configuration are found in figure \ref{fig:optical_design}. Since the smallest operational wavelength is $23.8$cm, the surface profile of the telescope mirrors should have an $rms$ error $\lesssim~15$ mm to allow for maximum efficiency. The final optical design is described in the work of \cite{BINGO_Abdalla:2020}.

\begin{figure}[h]
\begin{center}
\includegraphics[width=7.5cm]{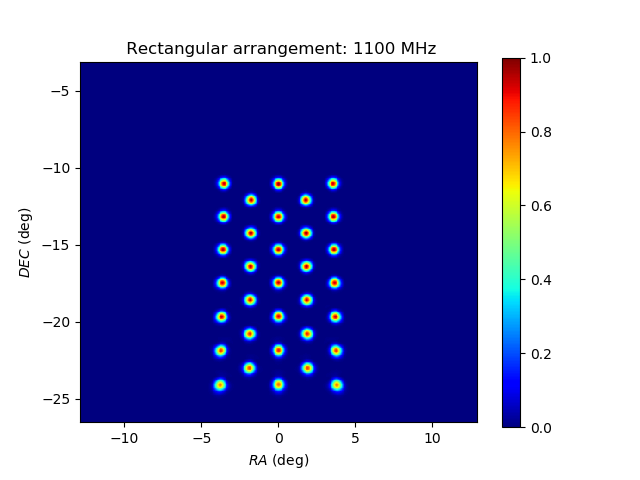}
\includegraphics[width=7.5cm]{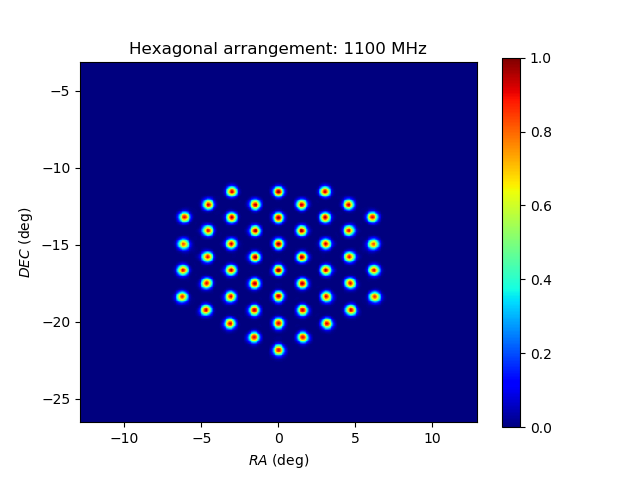}
\caption{Beam pattern response for the Stokes parameters as a function of celestial coordinates, for the focal plane arrangement of a rectangular horn arrangement configuration (left) and for a hexagonal horn arrangement configuration (right). The colour scale on the right refers to intensity (dB) \citep{BINGO_Abdalla:2020}.}
\label{fig:focalplane}
\end{center}
\end{figure}

\begin{figure}
\begin{center}
\includegraphics[width=9cm]{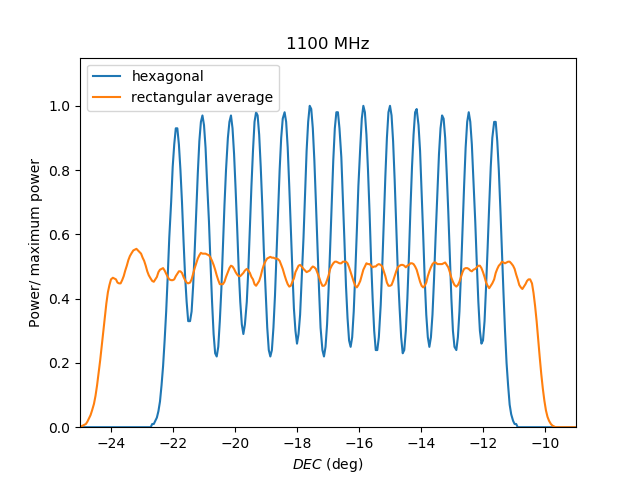}
\caption{Combined beam response for rectangular arrangement and hexagonal arrangement shown in figure \ref{fig:focalplane} at 1100 MHz \citep{BINGO_Abdalla:2020}.}
\label{fig:beam}
\end{center}
\end{figure}

%The gain of the telescope beams associated with the feed horns at the edge of the array are less than 1 dB lower compared with those from the horns at the centre; optical aberrations are slight, the edge beams being almost circular. 
Various arrangements for the horns in the focal plane were simulated in order to achieve an optimal sky coverage with the least distortion of the outer horns. All the optical analysis was performed with the GRASP package (TICRA - Reflector Antenna and EM Modelling Software)\footnote{\url{http://https://www.ticra.com/software/grasp/}}. The beam patterns shown in figure \ref{fig:focalplane} correspond to the best solutions for the horn arrangement in the focal plane. The corresponding beam response as a function of declination for both arrangements are shown in figure \ref{fig:beam}. 

The guiding principle in the design of BINGO is simplicity. All components should be as simple as possible to minimize costs. Moreover, since there will be no moving parts, design, operation and instrument modelling will also be simpler than doing the same tasks for a conventional telescope. Another key advantage of a simple design is that it is being built in a relatively small amount of time, allowing for results within a competitive science time window.  

\subsection{Receiver}

\begin{figure}[h]
\begin{center}
\includegraphics[width=10cm]{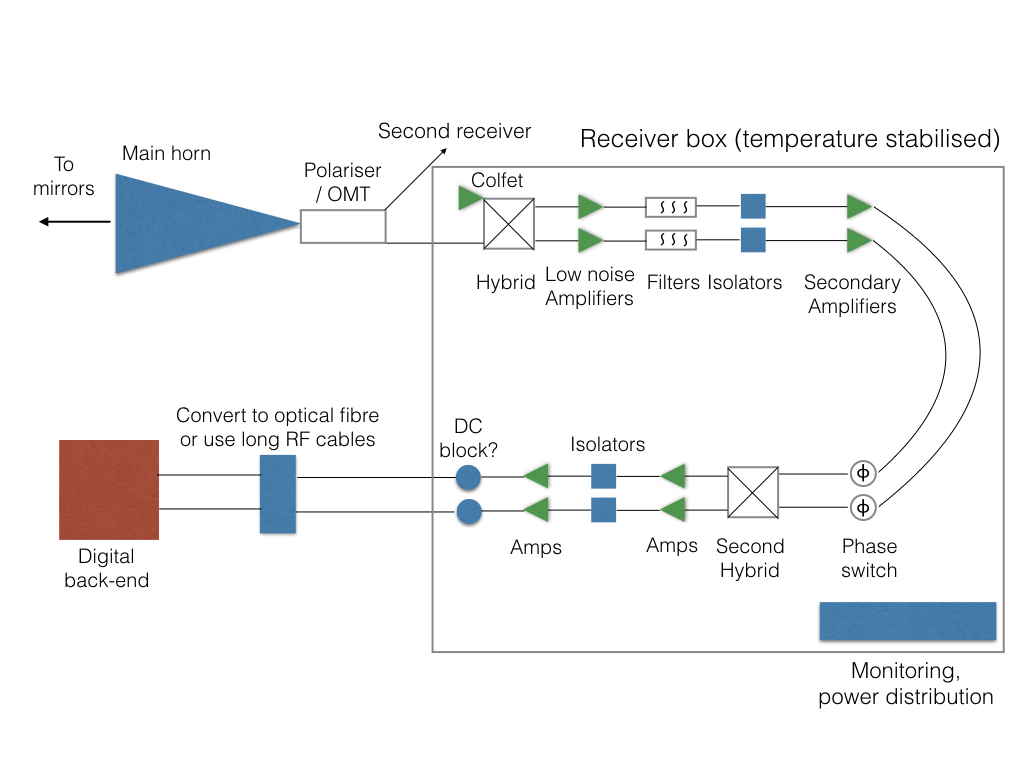}
\caption{Block diagram of the main components of the BINGO correlation receiver. The correlation is achieved by means of the hybrids, which in our case will be waveguide magic tees.}
\label{receiver1}
\end{center}
\end{figure}

Each receiver chain contains a correlation system, as shown in fig. \ref{receiver1}, operating with room temperature amplifiers. The receiver is expected to operate at temperature $T_{sys} \thicksim  70$ K in the frequency interval $980 - 1260$ MHz, at ambient temperature. A number of tests were performed at INPE's Division of Astrophysics, using two different low noise amplifiers (LNA). We tested a radiometer chain, with one LNA, one first stage filter and second stage LNAs. The testing was conducted in the laboratory, with no specific shielding in the amplifier structure. %The measured receiver noise temperature was $ T_R  = (58 \pm 6)$ K, taken as an upper limit, since the environment in which the measurements were made is permeated by RFI. 
Table \ref{comp_antena} shows the temperature for each component of the BINGO radio telescope. Thus, the system temperature is given by $ T_{\rm sys} = 69.9$ K. To estimate the sensitivity of the BINGO using this simple radiometer, two ratios are used, one for the minimum temperature variation that can be detected by the radiometer and another for the flow density that such temperature represents: 

\begin{equation}
\label{deltaTmin}
    \Delta T_{\rm min} = \frac{T_{\rm sys}}{\sqrt{B \tau}}
\end{equation}

\begin{equation}
\label{deltaSmin}
    S(\Delta T_{\rm min}) = \frac{2k}{A_{\rm eff}} \Delta T_{\rm min}.
\end{equation}

\noindent In eq. (\ref{deltaTmin}), B is the frequency operating bandwidth, $\tau $ the integration time, $k$ is the Boltzmann constant and $ A_{\rm eff} $ the effective area of the radio telescope. In this work, the bandwidth is $2.8 \times 10^8 $ Hz and the integration time is 1 s. The effective area used considers that 70 \% of the total area of the primary mirror receives electromagnetic waves. For BINGO, this area is $1134~ {\rm m}^2 $. The minimum temperature obtained in this case is $ \Delta T_ {\rm min} = 23$ mK and the corresponding flux, from eq. (\ref{deltaSmin}) is $ S (\Delta T_{min})  = 0.01$ Jy. 
After one year of observations, BINGO should achieve a $40'$ pixel noise of $84~\mu$K in a 9.33 MHz frequency channel, with the parameters described in Table \ref{tab:optics}. A more detailed discussion of the receiver tests can be found in the M.Sc. dissertation of \cite{Vieira:2020}. 

\begin{table}[h]
	\centering
	\scriptsize
	\caption{Temperature contributions of different components}
	\label{comp_antena}
	\begin{tabular}{|l|c|}
	\hline
	\textbf{Component} & \textbf{T (K)} \\
	\hline
	CMB & 2.7 \\
	Galactic/Extragalactic Signals & $\sim$~5,0 \\
	Ground spillover & 2.0 \\
	Atmosphere & 4.5 \\
	Parabolic Mirror & 1.5 \\
	Hyperbolic Mirror & 1.5 \\
	Horn (measured)                     & 6.8 \\
	Polarizer (measured)                & 9.1 \\
	Magic tee (measured)                & 2.8 \\
	Rectangular-to-coax waveguide (measured)	    & 9.1 \\
	LNA (phase matched)	                & 24.9 \\
	\hline
	Total & $\sim 69.9$ \\
	\hline
\end{tabular}
\end{table}

\subsection{Feedhorns and polariser}

BINGO will use specially designed conical corrugated horns to illuminate the secondary mirror of the telescope. These need to be corrugated in order to provide the required low sidelobes coupled with very good polarisation performance. Because of the large focal ratio needed to provide the required wide field of view, the horns have $\thicksim 1.9$m diameter and $\thicksim 4.3$m length. The electromagnetic design of such horns is well understood and the first prototype was successfuly developed and tested in the Laboratory of Integration and Tests at INPE. Insertion and return loss, as well as the beam profile, were measured for the frequency interval $900 - 1300$ MHz, in $25$ MHz steps, for vertical and horizontal polarisations. A summary of the main properties of the horn prototype is found in Tables \ref{tab:horn1} and \ref{tab:horn2} and the main results of the construction and testing of the feed horns and polariser are discussed in \cite{Wuensche:2019}. 

\addtolength{\tabcolsep}{-4pt}
\begin{table}[ht]
\caption{Feedhorn beam measurements. Left: horizontal polarisation. Right: vertical polarisation. Minimum rejection was not measured at all frequencies}
\scriptsize
\centering
\begin{tabular}{@{}c|cccc|cccc@{}}
\hline
Freq 		& Directivity 	& FWHM 	& Min. 	& Front to back	& Directivity 	& FWHM 	& Min. 	& Front to back \\
	 	& 	 		& 	 	& rejection  	& difference	& 	 		& 	 	& rejection  	& difference	\\
(MHz)	& (dB)		& (deg)	& (dB)		& (dB)		&  (dB)		& (deg)	& (dB)		& (dB)		\\	
\hline
\phantom{0}950 & 22.4 & 14.6 & -- & 47.0 & 22.7 & 13.4 & -- & 56.2\\
1000 & 23.1 & 13.8 & 22.5 & 51.7 & 23.0 & 13.5 & 20.5 & 64.7\\
1050 & 23.5 & 13.1 & 28.2 & 54.2 & 23.5 & 12.9 & 35.9 & 57.2\\
1100 & 23.9 & 12.4 & 25.6 & 58.2 & 23.9 & 12.4 & 30.5 & 56.2\\
1150 & 24.3 & 11.8 & 20.0 & 72.8 & 24.4 & 11.7 & 19.7 & 55.4\\
1200 & 24.7 & 11.1 & 26.4 & 57.2 & 24.6 & 11.1 & 26.0 & 63.6\\
1250 & 25.2 & 10.3 & -- & 56.1 & 24.9 & 10.6 & -- & 57.1\\
\hline
\end{tabular}
\label{tab:horn1}
\end{table}
\addtolength{\tabcolsep}{4pt}

\begin{table}[ht]
\caption{Horn measurements - summary}
\scriptsize
\centering
\begin{tabular}{@{}c|cc|cc@{}}
\hline
  \multicolumn{1}{c}{} &
  \multicolumn{2}{c}{Insertion loss} &
  \multicolumn{2}{c}{Return loss}  \\
\hline
            & measured              & predicted & measured  & predicted \\
\hline
Horn        &  $-$0.14\phantom{0}   &  --       & 26.50  & 28\phantom{.00} \\
Polariser   &  $-$0.12\phantom{0}   & $-$0.08   & 24.06  & 24.44    \\
WG5 to coax &  $-$0.075             & --        & $> 30$ & --   \\	
\hline
\end{tabular}
\label{tab:horn2}
\end{table}

\subsection{Mission simulations}

The assessment of the instrument performance and the reliability with which the cosmological signal can be extracted from the observed data is done through a \emph{Pipeline} developed inside the collaboration and used in previous works \citep{BigotSazy:2015,Harper:2018}. We briefly describe the data processing and component separation algorithm used to obtain the BINGO maps we use for our analysis. Details of the algorithm and data processing can be found in the works of \cite{BINGO_Liccardo:2020} and  \cite{BINGO_Fornazier:2020}. 

The \emph{Pipeline} takes input maps of different emission mechanisms, produced by theoretical models or by observations, as well as various instrument characteristics (such as such as the number of horns and their positioning in the focal plane, receiver system temperature, focal length, etc.) and contamination from the environment. \emph{Pipeline} outputs are time series which can be turned into maps that simulate the signal as measured by the instrument during a defined mission duration. 

To detect BAOs in the \textsc{Hi} signal we will need to remove the contributions from much brighter emissions coming from our Galaxy and extragalactic sources, the most relevant emissions at $\sim 1~{\rm GHz}$ being a combination of extragalactic point sources and diffuse Galactic synchrotron emission. %These two foregrounds combined contribute $\sim 1$ K rms at 1 GHz, while the 21cm signal fluctuations are $\sim$ 100 $\mu$K rms. Some very efficient methods of component separation are being tested for BINGO, so that we can recover the cosmological \textsc{Hi} component.

The \emph{Pipeline} operation makes use of a mission simulator, which processes all the input information described above, producing a time series (TOD), used to produce the maps at each frequency determined by the band and the number of channels. Currently, the set of generated maps  is processed by a component separation algorithm to recover the cosmological \textsc{Hi} signal. Future plans are that this procedure should be integrated in the \emph{Pipeline} body. 

We are currently simulating the instrument performance
% using a number of different optical arrangements produced by the collaboration \cite{BINGO_Abdalla:2020} 
and testing component separation methods, such as GNILC, developed for the Planck mission, following the work from \cite{Remazeilles:2011} and \cite{Olivari:2016}. 

A detailed description of the instrument performance and capabilities, as well as some discussion regarding the separation of components can be found in the work of  \cite{BINGO_Liccardo:2020}. A deeper analysis of the component separation process, especially regarding GNILC, and the construction of a non-gaussianity estimator module is presented in the work of \cite{BINGO_Fornazier:2020}.

\newpage

\section{Current status}
\label{sec:status}

BINGO is currently funded, with about 70\% of the funding already secured, mostly coming from Brazilian funding agencies (FAPESP, FINEP, MCTIC and CNPq) and a smaller amount coming from China, through purchase of components and electronic items for the instrument. 

There was a major project review in July 2019, with a committee of five scientists, including two radio astronomers and and one \textsc{Hi} IM specialist . The result of this review was very positive, with fourteen recommendations in total. Out of these fourteen, four of them require the close attention of the collaboration working groups and started being addressed immediately after the conclusion of the review. Namely, these four recommendations are:  1) the definition of the final receiver configuration to be used; 2) the final optical design; 3) the recommendation of adopting a full polarisation analysis (or a good motivation to avoid it); and 4) the preparation of an ``end-to-end'' pipeline to simulate the mission behaviour well in advance of the beginning of operations. We have solved items 1) and 3) and item 2) is being completed. We are presently working intensely in item 4). 

\section{Final remarks}

BINGO is a radio telescope designed to claim the first BAO detection, in the redshift interval $0.13 \leq z \leq 0.45$, at radio wavelengths around $1~{\rm GHz}$ and is currently being constructed in Para\'{\i}ba, Northeastern Brazil, counting on the local support and expertise of Campina Grande Federal University. The measured signal is produced by the redshifted 21cm HI line, detected through an IM survey covering $\sim~13\%$ of the sky and will probe the same redshift interval as the most important optical BAO surveys, but with different systematics. BINGO will also provide Galactic foreground maps in the frequency range 980 - 1260 MHz, with its sky coverage overlapping with a number of radio surveys. It will likely be the only radio instrument operating in its redshift range for a few years. 

BINGO will provide high quality data, covering a wide range of scientific areas from cosmology to Galactic science. In view of its observation strategy and with some adjustment in the digital backend, BINGO will also be capable to detect transient phenomena, such as pulsars and FRB. 

\section*{Author contributions} 
Carlos A. Wuensche and Elcio Abdalla are, respectively, the PI and the general coordinatior of the project. Filipe Abdalla, Luciano Barosi, Bin Wang, Amilcar Queiróz, Francisco Brito and Thyrso Villela belong to the management team, who are working in various aspects of logistics and construction to make BINGO a reality. Richard Battye, Clive Dickinson and Stuart Harper worked heavily in the initial stages of BINGO. Mike Peel, Ian Browne and Christian Monstein have a strong participation in the receiver construction, which is led by Carlos A. Wuensche; Mike Peel is the first author of the site selection paper and Telmo Machado has worked in many aspects of the receiver construction.  Christopher Radcliffe produced the filters and the simulations of the BINGO horns and front end. Jacques Delllabrouile, Mathieu Remazeilles, Larissa Santos, Karin Fornazier, Eduardo Mericia, Giancarlo de Gasperis and Vincenzo Liccardo are the contributors to the mission simulation and component separation aspects of the data analysis. André Costa, Ricardo Landim, Elisa Ferreira, Jianjun Zhang, Camila Novaes and Yin-Zhe Ma are the main developers of the cosmological forecast module in the data analysis effort. The optical design was led by Filipe Abdalla and Bruno Maffei, with the heavy involvement of Alessandro Marins and Pablo Motta, and the participation of João A. M. Barreto, Daniel Correia, Priscila Gutierrez, Milena M. M. Mendes, Juliana F. R. Santos, Carlos H. Otobone, Rafael M. G. Ribeiro, Karin Fornazier and Gustavo B. Silva, who ran the simulations of the optical system and produced the fits for the horn positions. The Fast Radio Bursts and the pulsar working group is led by Ricardo Landim, with the strong participation of Marcelo V. dos Santos, Frederico Vieira, Jordany Vieira, Chenxi Shan, Yu Sang and Xue Zhang. Rui An, Chang. Feng, Linfeng Xiao, Weiqiang Yang and Zenghao Zhu are working on different aspects of the cosmological parameter estimation effort.

\begin{acknowledgments}
C.A.W. acknowledges the CNPq grant 313597/2014-6 and FAPESP Thematic project grant 2014/07885-0. 
E.A. acknowledges FAPESP by the grant 2014/07885-0 and FAPESP and CNPq for various other grants. 
R.G.L. acknowledges CAPES (process 88881.162206/2017-01) and Alexander von Humboldt Foundation for the financial support. 
M.P. acknowledges funding from a FAPESP Young Investigator fellowship, grant 2015/19936-1. 
F.A.B acknowledges a CNPq/PRONEX/FAPESQ-PB (Grant no. 165/2018). 
C.A.W. and F.B.A. acknowledge the UKRI-FAPESP Visiting Professor Grant 2019/05687-0. 
The authors declare they have no conflicts of interest regarding the contents of this paper.
\end{acknowledgments}

%% To help institutions obtain information on the effectiveness of their 
%% telescopes the AAS Journals has created a group of keywords for telescope 
%% facilities.
%
%% Following the acknowledgments section, use the following syntax and the
%% \facility{} or \facilities{} macros to list the keywords of facilities used 
%% in the research for the paper.  Each keyword is check against the master 
%% list during copy editing.  Individual instruments can be provided in 
%% parentheses, after the keyword, but they are not verified.

\bibliography{BRICS_References2021}{}
\bibliographystyle{aasjournal}

%% This command is needed to show the entire author+affiliation list when
%% the collaboration and author truncation commands are used.  It has to
%% go at the end of the manuscript.
%\allauthors

%% Include this line if you are using the \added, \replaced, \deleted
%% commands to see a summary list of all changes at the end of the article.
%\listofchanges

\end{document}